\documentclass[10pt]{article}
\usepackage{fullpage,amssymb,amsmath}
\usepackage[dvips]{epsfig}

\def\##1{{\bf #1}}
\def\=#1{\underline{\underline{#1}}}

\def\+#1{\underline{\bf #1}}
\def\*#1{\underline{\underline{\bf #1}}}

\def\r#1{(\ref{#1})}
\def\l#1{\label{#1}}
\def\c#1{\cite{#1}}

\def\le{\left(}
\def\ri{\right)}
\def\les{\left[}
\def\ris{\right]}
\def\lec{\left\{}
\def\ric{\right\}}

\def\.{\mbox{ \tiny{$^\bullet$} }}

\def\epso{\epsilon_{\scriptscriptstyle 0}}

\def\muo{\mu_{\scriptscriptstyle 0}}

\def\ko{k_{\scriptscriptstyle 0}}
\def\co{c_{\scriptscriptstyle 0}}

\def\eps{\epsilon}

\begin{document}

\begin{center} {\bf {\LARGE Negative-- and positive--phase--velocity propagation in
 an isotropic chiral medium
moving at constant velocity}}
\end{center} \vskip 0.2cm

\noindent  {\bf Tom G. Mackay$^a$ and Akhlesh Lakhtakia$^b$} \vskip
0.2cm

\noindent {\sf $^a$ School of Mathematics\\
\noindent James Clerk Maxwell Building\\
\noindent University of Edinburgh\\
\noindent Edinburgh EH9 3JZ, United Kingdom\\
email: T.Mackay@ed.ac.uk} \vskip 0.4cm

\noindent {\sf $^b$ CATMAS~---~Computational \& Theoretical Materials Sciences Group \\
\noindent Department of Engineering Science \& Mechanics\\
\noindent 212 Earth \& Engineering Sciences Building\\
\noindent Pennsylvania State University, University Park, PA
16802--6812\\
email: akhlesh@psu.edu} \vskip 0.4cm

\vspace{15mm}

\begin{center} {\bf Abstract} \end{center}

Analysis of electromagnetic planewave propagation in a medium which is a
spatiotemporally homogeneous,
temporally nonlocal, isotropic, chiral medium in a co--moving frame
of reference shows that the medium is both spatially
and temporally nonlocal with respect to all non--co--moving
inertial frames of reference. Using the Lorentz transformations of electric and
magnetic fields, we show that plane waves which have
positive phase velocity in the co--moving frame of reference can
have negative phase velocity  in certain non--co--moving frames of
reference. Similarly, plane waves which have negative phase velocity
in the co--moving frame can have positive phase velocity  in certain
non--co--moving frames.

\vskip 0.2cm \noindent {\bf Keywords:} {\em Isotropic chiral medium,
Lorentz transformation, negative phase velocity, nonlocality}

\vspace{20mm}

\section{Introduction}

Analysis of planewave propagation in a frame of reference that is
uniformly moving with respect to the medium of propagation can lead
to the emergence and understanding of new phenomenons. In this
respect, the Minkowski constitutive relations, as widely described
in standard books \c{Chen,Kong}, are strictly appropriate to
instantaneously responding mediums only \c{BC}. For realistic
material mediums, recourse should be taken to the Lorentz
transformation of electromagnetic field phasors \c{Pappas,LW96}.

In an earlier study on plane waves in a medium that is spatiotemporally homogeneous,
temporally nonlocal,  isotropic and chiral in a co--moving frame
of reference, we reported that
planewave propagation with negative phase velocity (NPV) is possible with respect
to a
non--co--moving frame of reference, even though the medium does not support NPV
propagation in the co--moving frame \c{ML07}. That study  applies strictly
only at low translational speeds. In this paper,
we demonstrate  by means of an analysis
based on the Lorentz--transformed electromagnetic fields in the non--co-moving frame, that our
conclusion remains qualitatively valid for realistic mediums even at high translational
speeds.

\section{Planewave analysis}

We consider a spatiotemporally homogeneous,  spatially local, temporally
nonlocal, isotropic chiral
medium, characterized in the
 frequency domain by the Tellegen constitutive
 relations \c{Beltrami}
\begin{equation}
\left.
\begin{array}{l}
\#D' =   \epso \eps_r'\#E' + i \sqrt{\epso \muo} \xi' \#H'  \\ \vspace{-3mm} \\
\#B' =  - i \sqrt{\epso \muo} \xi' \#E' + \muo \mu_r'\#H'
\end{array}
\right\}, \l{Con_rel_p}
\end{equation}
in an inertial frame of reference $\Sigma'$. The
 relative permittivity $\eps_r'$, relative permeability $\mu_r'$ and
 chirality parameter $\xi'$
 are complex--valued functions of the angular frequency
 $\omega'$ if the medium is dissipative, and real--valued
 if it is nondissipative \cite[p. 71]{Chen};
 $\epso$ and $\muo$ are  the permittivity and permeability of
free space, respectively. The electromagnetic field phasors are related by
the Maxwell curl postulates as
\begin{equation} \l{MCP}
\left. \begin{array}{l}
 \nabla \times \#H' +  i \omega'  \#D' = \#0 \\ \vspace{-3mm} \\
\nabla \times \#E'  -   i \omega'  \#B' = \#0\end{array} \right\} .
\end{equation}

Our attention is focused on a  plane wave, described  by the
$\Sigma'$  phasors
\begin{equation}
\left.
\begin{array}{l}
\#E' =  \#E'_0 \exp \les i \le \#k' \. \#r' - \omega' t' \ri \ris \\ \vspace{-3mm} \\
\#H' = \#H'_0 \exp \les i \le  \#k' \. \#r' - \omega' t'  \ri \ris
\end{array}
\right\},  \l{pw_Sp}
\end{equation}
which propagates in the medium characterized by \r{Con_rel_p}, with
wavevector $\#k' = k' \, \hat{\#k}'$ and wavenumber $k' = \ko k'_r$.
There are four possibilities for the relative wavenumber: $k'_r \in
\lec k'_{r1},  k'_{r2}, k'_{r3}, k'_{r4} \ric$ where
\begin{equation}
\left.
\begin{array}{l}
k'_{r1} = \sqrt{\eps'_r \mu'_r} + \xi'\\
k'_{r2} = \sqrt{\eps'_r \mu'_r} - \xi'\\
k'_{r3} =  -k'_{r1} \\
k'_{r4} = -k'_{r4}
\end{array}
\right\}.
\end{equation}
 With respect to frame $\Sigma'$, the plane
wave is assumed to be uniform; i.e.,  $\hat{\#k}' \in \mathbb{R}^3$.

Suppose that the inertial frame $\Sigma'$ is moving at constant velocity
$\#v=v\hat{\#v}$ relative to another inertial frame
 $\Sigma$.
The electromagnetic field phasors in $\Sigma$ are related to those
in $\Sigma'$ by the Lorentz transformations \c{Pappas,LW96}
\begin{equation}
\left. \begin{array}{l} \#E = \displaystyle{\le \#E' \.\hat{\#v} \ri
\hat{\#v} +
 \gamma \,
\les \le \=I - \hat{\#v}\hat{\#v} \ri \. \#E' - \#v \times
\#B' \ris } \vspace{4pt}  \\
\#B = \displaystyle{\le \#B' \.\hat{\#v} \ri \hat{\#v} +
 \gamma \,
\les \le \=I - \hat{\#v}\hat{\#v} \ri \. \#B' + \frac{ \#v \times
\#E'}{\co^2} \ris  } \vspace{4pt} \\
\#H = \displaystyle{\le \#H' \.\hat{\#v} \ri \hat{\#v} +
 \gamma \,
\les \le \=I - \hat{\#v}\hat{\#v} \ri \. \#H' + \#v \times
\#D' \ris  } \vspace{4pt} \\
\#D = \displaystyle{\le \#D' \.\hat{\#v} \ri \hat{\#v} +
 \gamma \,
\les \le \=I - \hat{\#v}\hat{\#v} \ri \. \#D' - \frac{ \#v \times
\#H'}{\co^2} \ris }\end{array} \right\},  \l{Dp}
\end{equation}
where  $\=I = \hat{\#x}\,\hat{\#x}+
\hat{\#y}\,\hat{\#y}+\hat{\#z}\,\hat{\#z}$ is the 3$\times$3
identity dyadic, $ \gamma = 1/\sqrt{1-\beta^2}$ and the relative translational
speed $\beta = v / \co$, with $\co=1/\sqrt{\epso\muo}$ being
 the speed of light in free space.
In terms of the $\Sigma$ phasors, the plane wave is described by
\begin{equation}
\left.
\begin{array}{l}
\#E = \#E_0 \exp \les i \le \#k \. \#r - \omega t \ri \ris \\ \vspace{-3mm} \\
\#H = \#H_0 \exp \les i \le  \#k \. \#r - \omega t  \ri \ris
\end{array}
\right\}.   \l{pw_S}
\end{equation}
The phasor amplitude vectors $\lec \#E_0, \#H_0 \ric$ and $\lec
\#E'_0, \#H'_0 \ric$ are related via the transformations \r{Dp},
whereas \c{Pappas}
\begin{eqnarray}
&&\#r = \les \, \=I  + \le \gamma - 1 \ri \hat{\#v}\,\hat{\#v} \ris
\.
\#r' + \gamma \, \#v t', \\
&& t = \gamma \le t' + \frac{\#v \. \#r'}{\co^2} \ri,\\
&& \#k = \gamma \le \#k' \. \hat{\#v} + \frac{\omega' v}{\co^2} \ri
\hat{\#v} + \le \, \=I - \hat{\#v} \, \hat{\#v} \ri \. \#k', \l{kT}\\
&& \omega = \gamma \le \omega' + \#k' \. \#v \ri. \l{wT}
\end{eqnarray}

Since $k' \in \mathbb{C}$ for a dissipative medium, we have from
\r{kT} that $\#k = k_R \hat{\#k}_R + i k_I \hat{\#k}_I$ with $k_R\in
\mathbb{R}$, $k_I \in \mathbb{R}$,  $\hat{\#k}_R\in \mathbb{R}^3$,
and $ \hat{\#k}_I \in \mathbb{R}^3$, but $\hat{\#k}_R \neq
\hat{\#k}_I$ in general; i.e., the plane wave is generally
nonuniform with respect to $\Sigma$. Similarly, from \r{wT} we have
that $\omega = \omega_R + i \omega_I$ with $\omega_R\in \mathbb{R}$
and $ \omega_I \in \mathbb{R}$. Expressing the $\Sigma$ phasors as
\begin{equation}
\left.
\begin{array}{l}
\#E = \lec  \#E_0 \, \exp \les - \le \#k_I \. \#r - \omega_I t \ri
\ris \ric
\exp \les i \le \#k_R \. \#r - \omega_R t \ri \ris \\ \vspace{-3mm} \\
\#H = \lec  \#H_0 \, \exp \les - \le \#k_I \. \#r - \omega_I t \ri
\ris \ric  \exp \les i \le  \#k_R \. \#r - \omega_R t  \ri \ris
\end{array}
\right\},  \l{pw_Sd}
\end{equation}
we note that that the periodic propagation of phase is governed by
$\#k_R$ and  $\omega_R$, whereas attenuation or growth of the wave
amplitude is governed by $\#k_I$ and  $\omega_I$. By writing the
phasor amplitudes as $\#E_0 = \#E_{0R} + i \#E_{0I}$ and $\#H_0 =
\#H_{0R} + i \#H_{0I}$,
the corresponding cycle--averaged Poynting vector may be expressed
as
\begin{eqnarray}
\#P &=&   \exp \le - 2 \#k_I \. \#r  \ri \frac{\omega_R}{2\pi}\,
\times \nonumber
\\
&&\Bigg[   \#E_{0R} \times \#H_{0R}   \int^{t_0 + \frac{2 \pi}{
\omega_R}}_{t_0} \cos^2 \le \#k_R \. \#r - \omega_R t \ri \, \exp
\le 2 \omega_I t \ri \; dt \nonumber \\ &&
 -  \le  \#E_{0I} \times \#H_{0R} +  \#E_{0R} \times \#H_{0I} \ri\times
 \nonumber
 \\
 &&\qquad
   \int^{t_0 + \frac{2 \pi}{
\omega_R}}_{t_0} \cos \le \#k_R \. \#r - \omega_R t \ri \,\sin \le
\#k_R \. \#r - \omega_R t \ri \, \exp \le 2 \omega_I t \ri \; dt
\nonumber \\ && +  \#E_{0I} \times \#H_{0I}   \int^{t_0 + \frac{2
\pi}{ \omega_R}}_{t_0} \sin^2 \le \#k_R \. \#r - \omega_R t \ri \,
\exp \le 2 \omega_I t \ri \; dt  \Bigg]
\end{eqnarray}
for a cycle beginning at time $t=t_0$. The phase velocity is given
by
\begin{equation}
\#v_p = \frac{\omega_R}{k_R}\, \hat{ \#k}_R.
\end{equation}
Whether the plane wave has positive phase velocity (PPV) or negative
phase velocity (NPV) in the reference frame $\Sigma$ is determined
by the sign of $\#v_p \.  \#P $: positive for PPV and negative for
NPV. Criterions for determining whether the phase velocity is
positive or negative with respect to the  reference frame $\Sigma'$
are presented elsewhere \c{M05,ML07_MOTL}.

\section{Numerical results and discussion}

For the sake of illustration, let us consider the cycle--averaged
Poynting vector evaluated at the point $\#r = \#0$ with the temporal
averaging starting from  $t_0 = 0$; i.e.,
\begin{eqnarray}
 \left. \#P \right|_{\#r= \#0}  &=&    \frac{\omega_R }{8 \pi \omega_I | \omega |^2} \les \exp \le \frac{ 4 \pi
\omega_I}{  \omega_R} \ri - 1 \ris \Big[  \le | \omega |^2 +
\omega^2_I \ri   \#E_{0R}
\times \#H_{0R}  \nonumber \\
&&
 - \omega_R \omega_I   \le  \#E_{0I} \times \#H_{0R} +  \#E_{0R} \times \#H_{0I} \ri
    + \omega^2_R \, \#E_{0I} \times
\#H_{0I} \Big].
\end{eqnarray}
Without loss of generality, let us assume that the plane wave
propagates along the $z'$ Cartesian axis; i.e., $\hat{\#k}' =
\hat{\#z}'$. It follows then from the Maxwell curl postulates
\r{MCP} that  $\#E'_0$  lies in the $x'y'$ plane with
\begin{equation}
\left.
\begin{array}{lcr}
\hat{\#y}' \. \#E'_0 = i \hat{\#x}' \. \#E'_0 & \mbox{for} & k'_r =
k'_{r1}, k'_{r4} \\
\hat{\#y}' \. \#E'_0 = -i \hat{\#x}' \. \#E'_0 & \mbox{for} & k'_r =
k'_{r2}, k'_{r3}
\end{array}
\right\}.
\end{equation}
Further, we take  the velocity $\#v$ to lie in the $x'z'$ Cartesian
plane as per
\begin{equation}
\hat{\#v} = \hat{\#x}' \sin \theta+\hat{\#z}' \cos
\theta\,.
\end{equation}

\begin{figure}[!ht]
\centering \psfull \epsfig{file=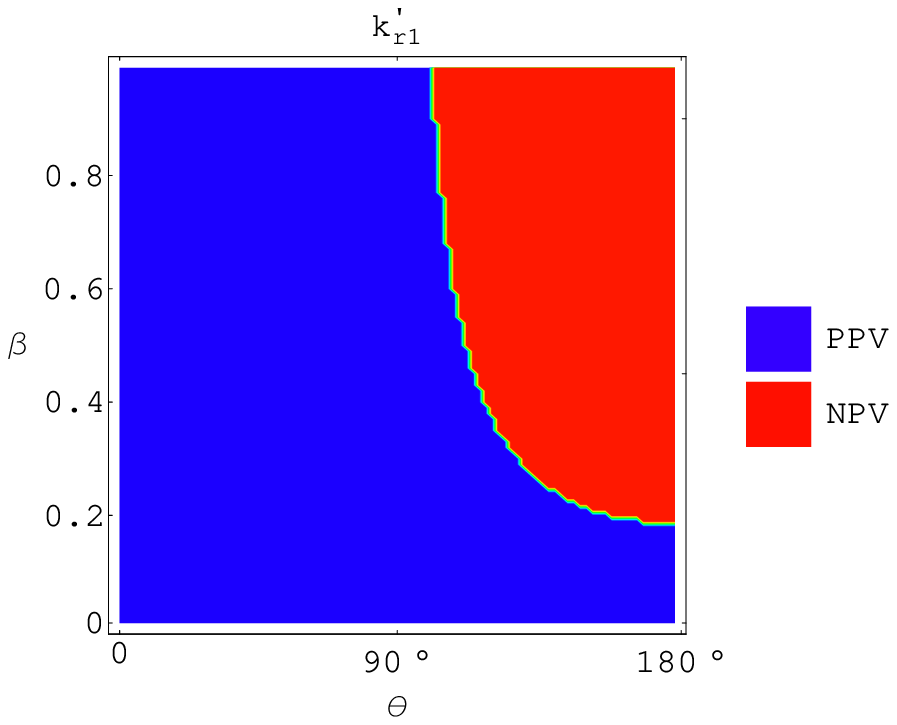,width=3.0in} \hfill
\epsfig{file=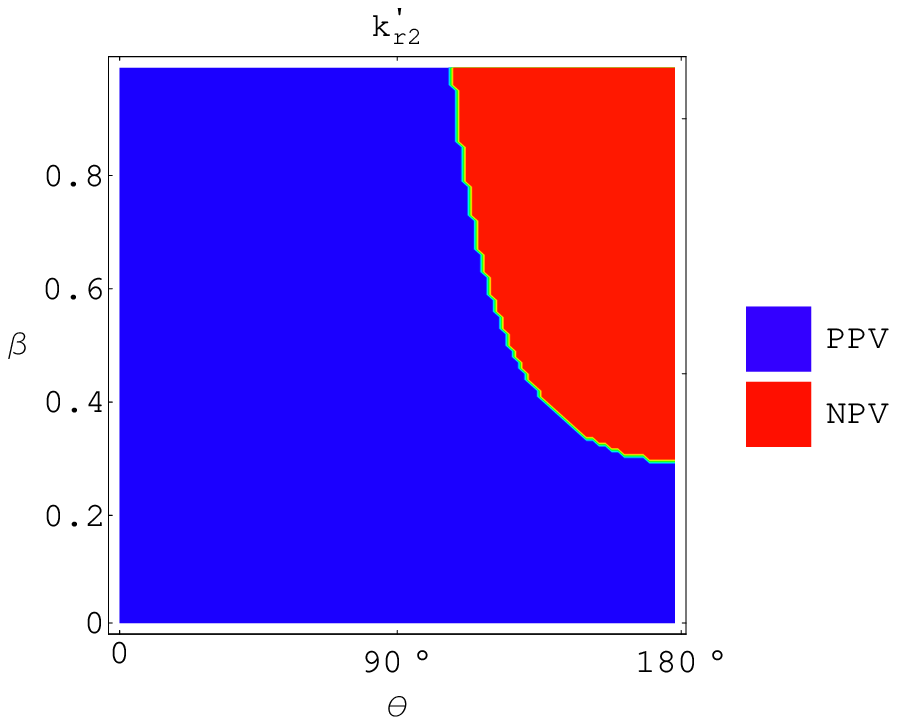,width=3.0in}
\\ \vspace{2mm}
\epsfig{file=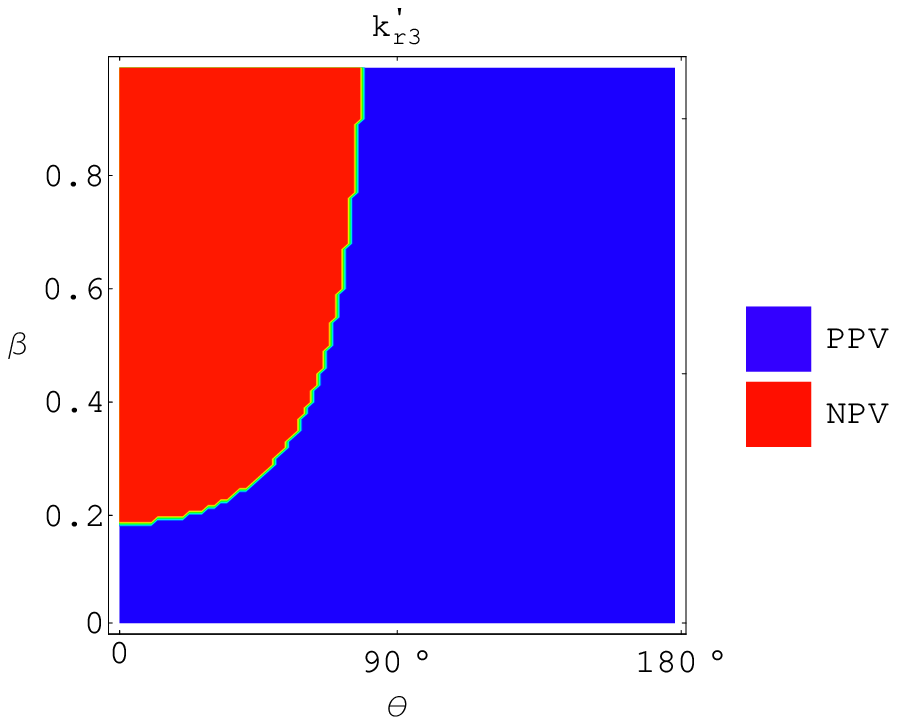,width=3.0in} \hfill
\epsfig{file=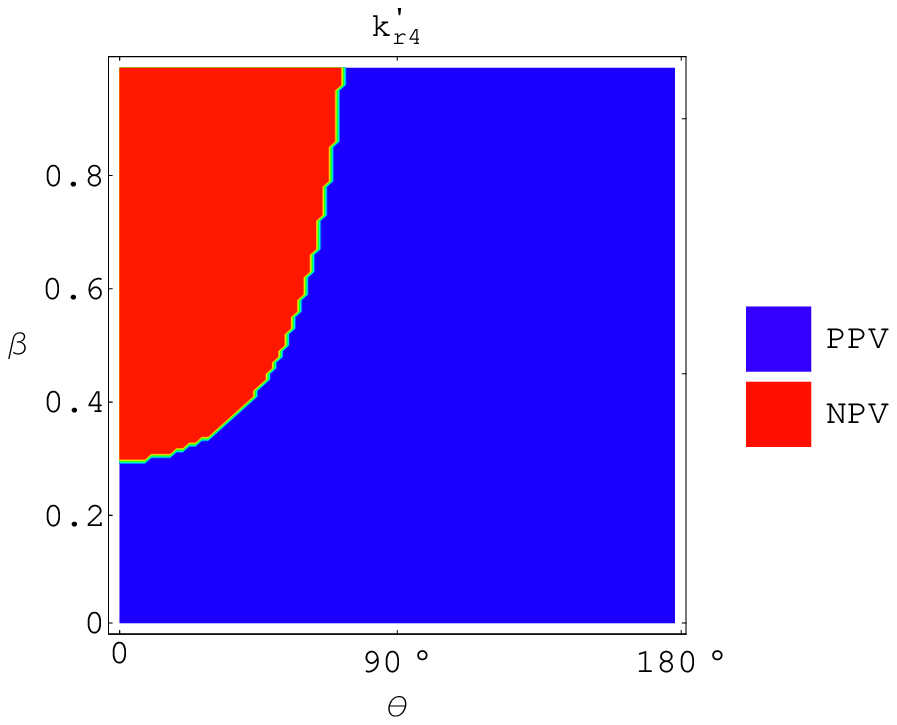,width=3.0in}
  \caption{\label{fig1}
  The distribution of positive phase velocity (PPV) and negative
  phase velocity (NPV) in the $\Sigma$ reference frame,
in relation to $\beta \in [ 0,1)$ and $\theta \in [ 0^\circ,
180^\circ)$, for $k'_r \in \lec k'_{r1},  k'_{r2}, k'_{r3}, k'_{r4}
\ric$. Here, $\eps'_r = 6.5 + 1.5i$, $\xi' = 1 + 0.2 i$
 and $\mu'_r = 3 + 0.5i$.
 }
\end{figure}

In Figure~\ref{fig1} the distributions of PPV and NPV in the
reference frame $\Sigma$ for the dissipative  scenario wherein
$\eps'_r = 6.5 + 1.5i$, $\xi' = 1 + 0.2 i$
 and $\mu'_r = 3 + 0.5i$ are displayed for $k'_r \in
\lec k'_{r1},  k'_{r2}, k'_{r3}, k'_{r4} \ric$. Clearly, for all
four values of $k'_r$, propagation is of the PPV type for $\beta =
0$ (i.e., with respect to the $\Sigma'$ frame). As the relative
translational speed increases, the phase velocity of the plane waves
corresponding to $k'_{r1}$ and $k'_{r2}$ eventually becomes negative
provided that $ \pi/2 < \theta < \pi$. In contrast the plane waves
corresponding to $k'_{r3}$ and $k'_{r4}$  have NPV at sufficiently
large values of $\beta$ provided that $ 0 < \theta < \pi/2$.

\begin{figure}[!ht]
\centering \psfull \epsfig{file=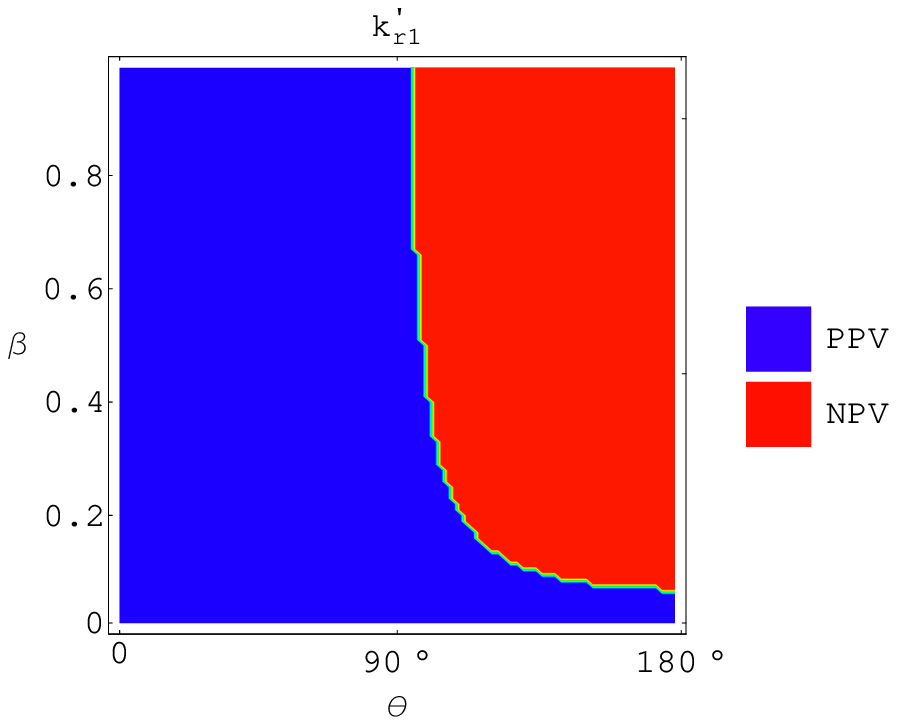,width=3.0in} \hfill
\epsfig{file=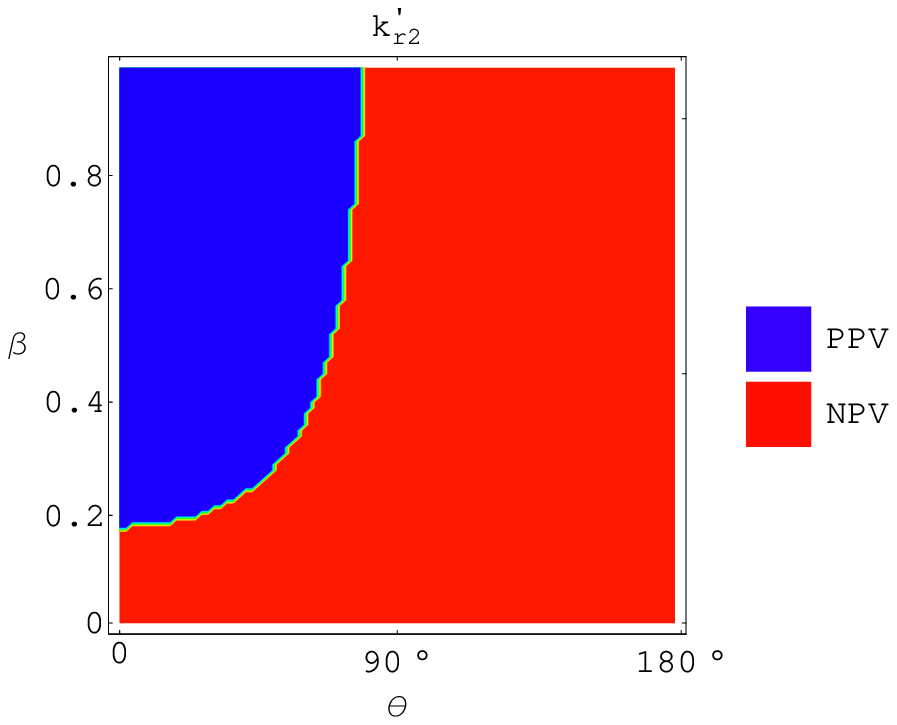,width=3.0in}
\\ \vspace{2mm}
\epsfig{file=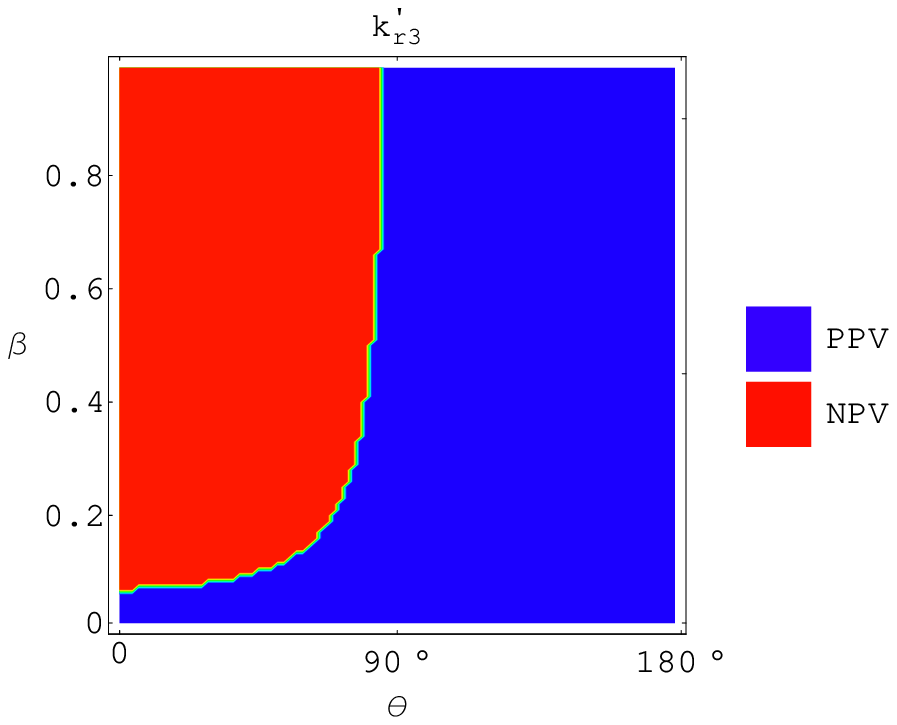,width=3.0in} \hfill
\epsfig{file=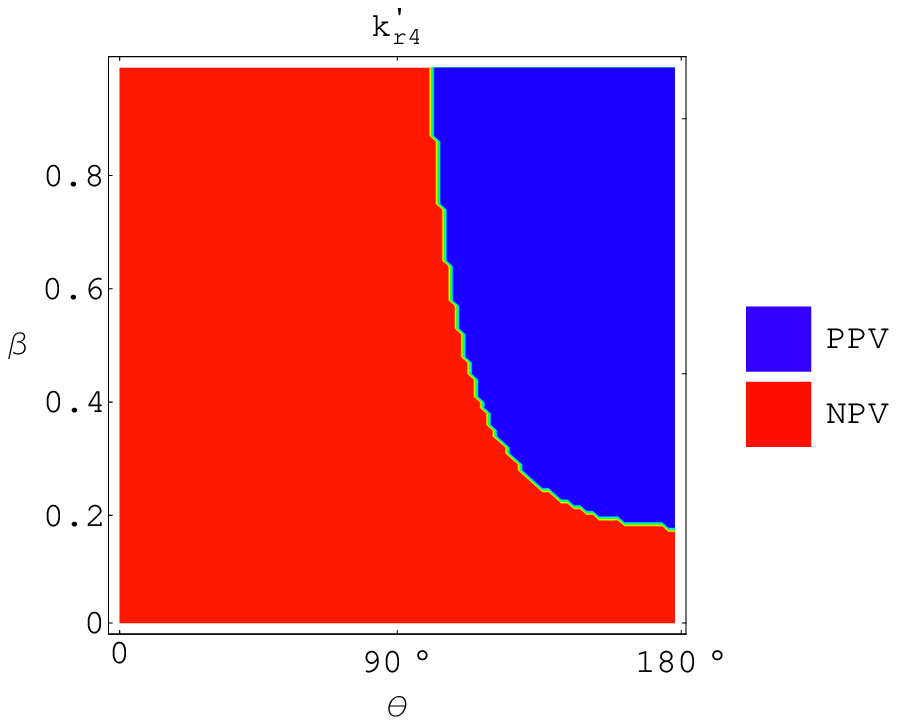,width=3.0in}
  \caption{\label{fig2}
  As Figure~\ref{fig1} but with $\xi' = 10 + 2i$.
 }
\end{figure}

Now, let us look at the scenario where the isotropic chiral medium
supports NPV propagation in the co--moving reference frame. This
situation arises when $| \mbox{Re} \, \lec \xi' \ric | > \mbox{Re}\,
\lec \sqrt{\eps'_r} \sqrt{ \mu'_r } \ric$, for example
\c{M05,ML07_MOTL}. We take $\eps'_r = 6.5 + 1.5i$, $\xi' = 10 + 2 i$
 and $\mu'_r = 3 + 0.5i$. The corresponding distributions of NPV and
 PPV are mapped against $\beta$ and $\theta$ in Figure~\ref{fig2}.
We see that the plane waves corresponding to relative wavenumbers
$k'_{r1}$ and $k'_{r3}$  have NPV when $\beta$ is small but have PPV
when $\beta$ is sufficiently large. On the other hand, the plane
waves corresponding to relative wavenumbers  $k'_{r2}$ and $k'_{r4}$
have PPV when $\beta$ is small but have NPV when $\beta$ is
sufficiently large.

To conclude, by means of the Lorentz--transformed electromagnetic fields,
we have
demonstrated that a plane wave with PPV in an isotropic chiral
medium can have NPV when observed from a no--co--moving inertial reference
frame. Similarly a NPV plane wave in the co--moving frame can be PPV
from a non--co--moving frame.

\vspace{10mm}

 \noindent {\bf Acknowledgements} The authors are most
grateful to Professor I.M. Besieris (Virginia Polytechnic Institute
and State University) for a discussion
on the Minkowski constitutive relations for realistic mediums. TGM
is supported by a \emph{Royal Society of Edinburgh/Scottish
Executive Support Research Fellowship}.

\newpage

\newpage

\end{document}